\documentclass[12pt]{article}
\usepackage{latexsym,graphicx,color}
\usepackage{colordvi}
\usepackage[portrait,margin=1in]{geometry}

\newcommand{\be}{\begin{equation}}
\newcommand{\ee}{\end{equation}}

\long\def\comment#1{}
\def\pd{\partial}

\def\pd{\partial}

\def\Alf{Alfv\'en }
\def\rfig#1{Fig.\ref{fig:#1}}
\def\rsec#1{Sec.\ref{sec:#1}}

\def\req#1{(\ref{eq:#1})}

\def\macname{hankrs2}
\def\macname{hank_strauss}
\def\fignf12{./pix}

\def\figdircc1{/Users/{\macname}/Documents/progs/m3dc1/plots}

\def\figdird3d{/Users/{\macname}/Documents/progs/m3d/d3d-lock}

\def\figdirp5{/Users/{\macname}/Documents/papers/disruption/rwtm/deltap/paper5}

\def\figdir{.}

\begin{document}
\begin{center}
{\bf\Large Prevention of resistive wall tearing mode major disruptions with feedback} \\
{H. R. Strauss$^{1 ,} $ \footnote{Author to whom correspondence should be 
addressed: hank@hrsfusion.com}}  \\
{$^1$ HRS Fusion, West Orange, NJ 07052 } 
\end{center}

\vspace{.5cm}

\abstract{
Resistive wall tearing modes (RWTM) can cause major disruptions. 
A signature of RWTMs is that the rational surface is
sufficiently close to the wall.
For $(m,n) = (2,1)$ modes, at
  normalized minor radius $\rho = 0.75$,
 the value of $q$  is
$q_{75} < 2.$ 
This is confirmed in simulations and theory and
in a DIII-D locked mode disruption database. 
The $q_{75} < 2$ criterion is valid at
high $\beta$ as well as at low $\beta.$ 
A very important feature of RWTMs is that they produce major disruptions 
only when the  $q_{75} < 2$ criterion is  satisfied. If it is not satisfied, 
or if the wall is ideally
conducting, then the mode does not produce a major disruption,
although it can produce a minor disruption. Feedback, or rotation of the
mode at the wall by complex feedback, can emulate an ideal wall, 
preventing major disruptions.
The $q_{75}$ criterion is analyzed in linear simulations, and a
simple geometric model is given. 
}

\section{Introduction} \label{sec:intro}

Resistive wall tearing modes (RWTM) can cause cause major disruptions.
This is based on evidence from theory, simulations, and experimental data
\cite{jet21,iter21,d3d22,mst23,model,nf-iaea24}.
For example, DIII-D locked mode shot 154576 \cite{sweeney} experienced 
a major disruption.
Linear simulations  \cite{d3d22} found the reconstructed equilibrium was stable
with an ideal wall. 
and found a scaling of the linear growth rate with the wall penetration time. 
 Nonlinear simulations found  a complete thermal quench, and  agreement with
the experimental thermal quench (TQ) time and the amplitude of the perturbed magnetic
field. 

A signature of RWTMs is that the rational surface is
sufficiently close to the wall. 
For $(m,n) = (2,1)$ modes, the rational surface radius of the $q=2$ surface, 
normalized to the plasma minor radius, is $\rho_{q2} > 0.75.$
This can also be expressed as the value of $q$ at $\rho = 0.75$, 
$q_{75} < 2.$ This is confirmed in simulations and theory. Experimentally, it is the
disruption criterion in a DIII-D locked mode disruption database. 
The importance of mode locking and disruption precursors is discussed.
The $q_{75} < 2$ criterion is valid at
high $\beta$ as well as at low $\beta.$ This   is verified experimentally as well as
in simulations.

A very important feature of RWTMs is that they produce major disruptions when the
$q_{75} < 2$ or $\rho_{q2} > 0.75$ criterion is  satisfied. If the wall is ideally
conducting, then the mode becomes a tearing mode and does not produce a major disruption,
although it can produce a minor disruption. Feedback, or rotation of the
mode at the wall by complex feedback, can emulate an ideal wall. This implies that
RWTMs can be made to act like tearing modes with an ideal wall, and only produce
minor disruptions. This is verified experimentally and in simulations at low and
high $\beta.$

The $q_{75}$ criterion for a RWTM implies that the $q = 2$ rational surface is sufficiently
close to the wall to interact with it. This is analyzed in a linear model, and a
simple geometric model is given of the wall interaction criterion. 
The criterion depends weakly on the ratio of $\rho_{q2} / \rho_w,$ where
$\rho_w$ is the wall radius normalized to plasma radius.  For $\rho_w > 1.5,$ the
wall is too far away for a RWTM and feedback stabilization is not possible. The
criterion is also obtained for general $(m,n),$ and requires the rational surface
to be closer to the wall. 
 
The outline of the paper is as follows. The domain of instability of RWTMs
in the $(q_{75},\beta)$ plane is presented qualitatively in \rsec{param}. Also
shown is the $\rho_{q2} > 0.75$ database of DIII-D locked mode disruptions. 
The relevance of mode locking, precursors, edge cooling, and current
contraction are discussed.
In \rsec{lowbeta}, simulations are presented of a sequence of equilibria in which
major disruptions occur with a resistive wall, when $q_{75} < 2,$ otherwise
a minor disruption occurs. If the wall is ideal, 
only a minor disruption occurs. In a particular example, it is demonstrated that 
 feedback or wall
rotation give a  similar result as an ideal wall. The amplitude of the 
non axisymmetric $n > 1$   perturbations is much larger when the wall is resistive, 
in comparison
with an ideal wall, feedback or rotating wall. 
The computational model used for feedback and  rotating wall is discussed in
\rsec{feedback}. Finite $\beta$ experimental results in NSXT are presented in 
\rsec{nstx}. The experiment shows a feedback limited tearing mode, with
$\rho_{q2} \approx 0.75,$ evidently close enough to the wall  to be feedback controlled. 
\rsec{nstxsim} shows simulations based on an NSTX intermediate $\beta_N$ equilibrium. 
For a resistive wall, a major disruption occurs. With an ideal wall, feedback, or
rotating wall, only minor disruptions occur. The 
 amplitude of the $n > 1$  perturbations is much larger for a resistive wall than
for ideal wall, feedback, or rotating wall, as in \rsec{lowbeta}. 
The reason for the $q_{75}$ criterion is analyzed in \rsec{q75}. The
critical value of $\rho_{q2} / \rho_w$ is obtained from linear stability of model
equilibria. It is in good agreement with a simple geometric model. The $\rho_{q2} > 0.75$
criterion occurs for $\rho_w = 1.2,$ as in DIII-D, the model equilibria of \rsec{lowbeta},
and NSTX. The critical $\rho_{q2}$ depends weakly on $\rho_w,$ 
for $\rho_w \le 1.5.$  A summary is provided in \rsec{summary}.


\section{RWTM parameter space} \label{sec:param}

\begin{figure}[h]
\begin{center}
\includegraphics[height=4.5cm]{\figdir/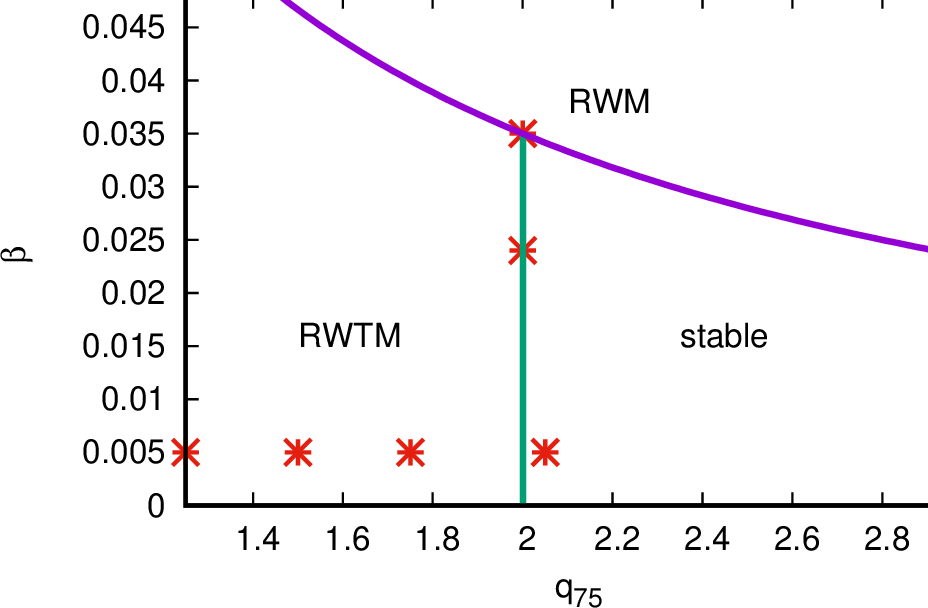}(a)
\vspace{.5cm}\hspace{.5cm}
\includegraphics[height=4.5cm]{\figdir/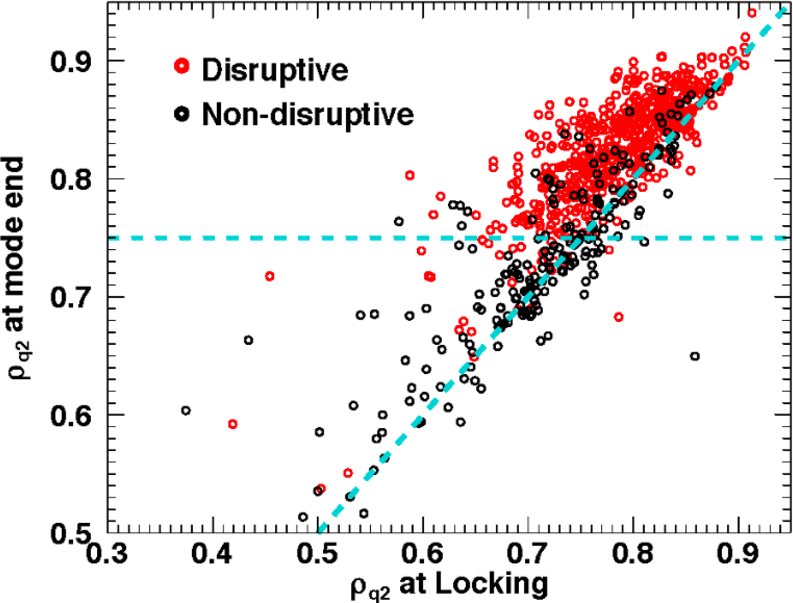}(b)
\end{center}
\caption{\it (a) Schematic diagram of RWM and RWTM stability in $(q_{75},\beta)$ space.
(b) Disruptivity in a DIII-D locked mode disruption database. Reproduced from \cite{sweeney2017}}
\label{fig:q75beta}
\end{figure}

The expression $\rho_{q2} = 0.75,$ can 
be written as $q_{75} = 2,$ where $q_{75} = q(\rho_{q2} = .75),$
which is useful to represent the RWTM unstable parameter space.  
\rfig{q75beta}(a) gives a  schematic
 parameter space $(q_{75},\beta)$
of RWM and RWTMs. The RWTM is unstable for  $q_{75} \le  2.$
The RWM beta limit is approximately the Troyon \cite{troyon}  no wall
limit $\beta_{N}$. The RWTM is unstable below the RWM limit \cite{betti,villamora}.
The labeled points correspond to  low and high  $\beta$ examples  in 
\rsec{lowbeta}, \rsec{nstx}, and \rsec{nstxsim}.
Both low and high $\beta$ RWTM and RWM can be limited to minor disruptions by feedback,
rotation, or an ideal wall.
Locking with resistive wall and without  feedback or rotation  allows a major disruption.

\vspace{.5cm}
\rfig{q75beta}(b) shows a  database of
 DIII-D disruptivity \cite{sweeney2017}  which depends on $\rho_{q2}.$ 
The onset is $\rho_{q2} = .75$ or $q_{75} = 2.$ 
Nearly all disruptions
occur for $\rho_{q2} > 0.75.$ 
The disruptions occur for locked modes.
Mode locking means that toroidal rotation stops,   
destabilizing  tearing modes \cite{sweeney,gerasimov2020}.
Sheared rotation stabilizes tearing modes  \cite{coelho,wang,drift},
 including RWTMs \cite{finn95}. 

\rfig{q75beta}(b) also shows that $\rho_{q2}$ tends to increase
between mode locking and the disruption.
This could be explained 
by edge cooling, which produces  current
contraction \cite{schuller,wesson}, 
and can  cause $\rho_{q2}$ to increase.
A  current contraction model \cite{model}  is discussed in \rsec{q75}.
Current  contraction is caused by edge cooling, which in turn can have several causes. 
One possible cause is overlapping tearing modes in the edge region, called a $T_{e,q2}$ collapse
\cite{sweeney}.  
Another possibility is  resistive ballooning turbulence, proposed as an explanation of
the Greenwald density limit \cite{ricci}.
Another possible cause of edge cooling  is impurity radiation \cite{pucella}. 
The impurity content might be raised by increasing the plasma density to the
Greenwald limit \cite{gates}. 
The impurities might be
introduced purposefully, as in massive gas injection \cite{izzo}, or accidentally,
as UFOs,  pieces of plasma facing tiles falling into the plasma. 
All these have been called causes of disruptions, but they are really precursors,
cooling the edge and destabilizing a RWTM.

\begin{figure}[h]
\begin{center}
\includegraphics[height=3.8cm]{\figdir/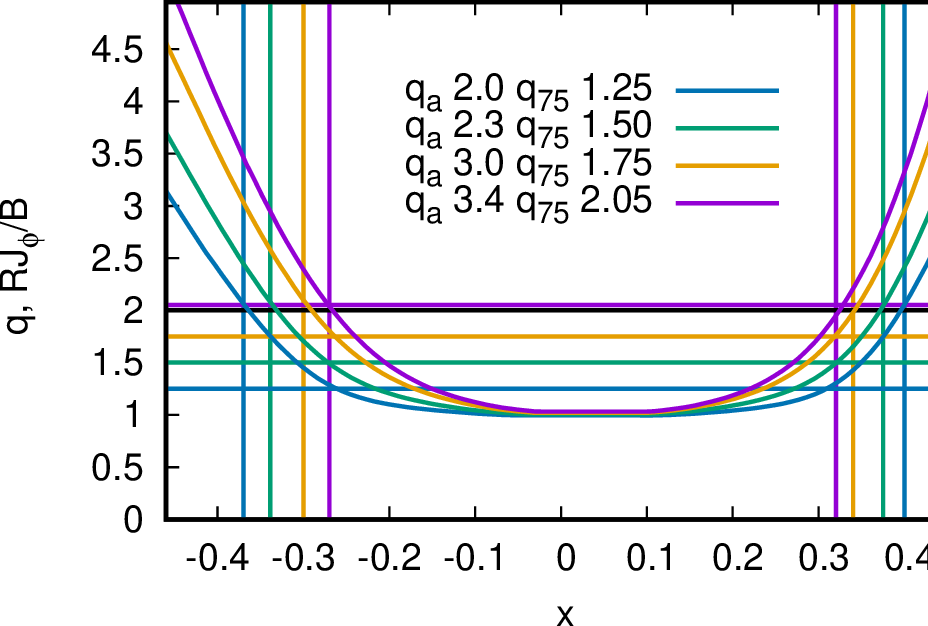}(a) 
\includegraphics[height=3.8cm]{\figdir/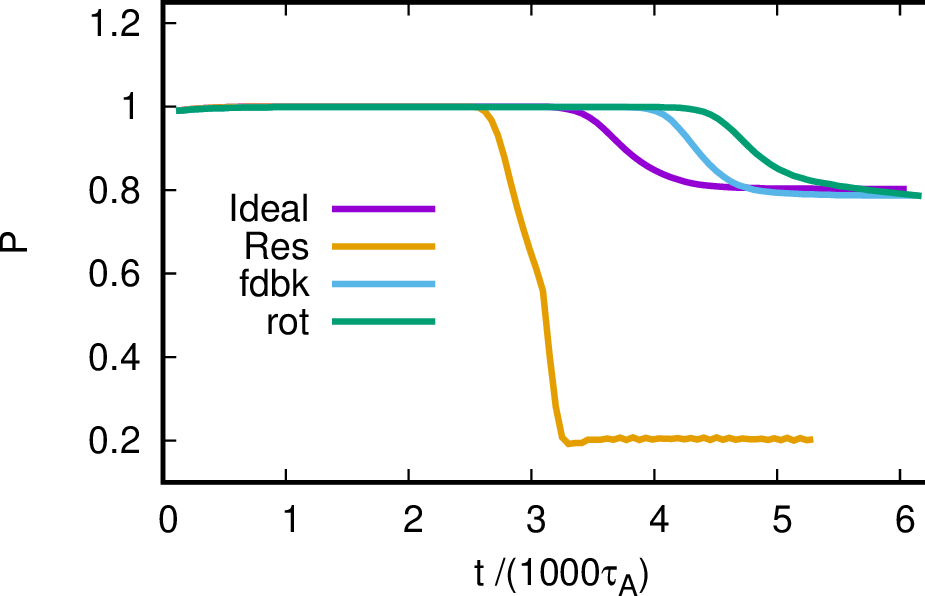}(b)
\end{center}
\caption{\it (a) $q$ profiles of model equilibria  as a function of major radius $x = R - R_0$ for
model equilibria with $q_a = 2, 2.3, 3, 3.4.$  All but $q_a = 3.4$ have
$q_{75} < 2.$
(b) time histories of case $q_a = 3$ with ideal, resistive,
feedback, and rotating wall boundary conditions. In all but the resistive wall case,
only a minor disruption occurs.
 } 
\label{fig:qmst} 
\end{figure}

\section{Low $\beta$ RWTM disruptions} \label{sec:lowbeta}

Simulations were performed with M3D \cite{m3d} for a sequence of modified
MST equilibria \cite{model,nf-iaea24}. Here the results are extended by
including simulations with feedback and wall rotation, as discussed in \rsec{feedback}.
The simulations had parameters: Lundquist number $S = 10^5,$ wall Lundquist
number $S_w = \tau_w / \tau_A = 10^3,$ where $\tau_w$ is the wall penetration
time and $\tau_A$ is the \Alf time, and parallel thermal conductivity
$\kappa_\parallel = 10 R^2 / \tau_A.$ The simulation had $16$ toroidal planes.   
\rfig{qmst} shows $q(x)$ profiles for a sequence of modified MST
equilibria \cite{model} with $\rho_w = 1.2.$ The profiles have $q_0 = 1$ and edge
$q_a = 2, 2.3, 3, 3.4.$ For $q_a \le 3,$ $q_{75} < 2,$ so the equilibria
are unstable to RWTMs. The case $q_a = 3.4$ is RWTM marginally stable.
Nonlinear simulations  with M3D \cite{m3d} were initialized with  these
equilibria. It was shown \cite{model} that with an ideal wall, all the equilibria
are unstable only to minor disruptions. For $q_a \le 3,$ $q_{75} < 2,$ with a 
resistive wall, major disruptions occur. If $q_{75} > 2$, and the wall is
resistive, the disruption is minor. The particular case $q_a = 3$, $q_{75} = 1.75$ 
is considered in more detail.
\rfig{qmst} (b) shows time histories of total pressure $P$ for the case $q_a = 3,$
with ideal wall, resistive wall, feedback, and wall rotation.
A major disruption occurs for a resistive wall. All the other boundary 
conditions give only minor disruptions.
\begin{figure}[h] \begin{center}
\includegraphics[height=2.5cm]{\figdir/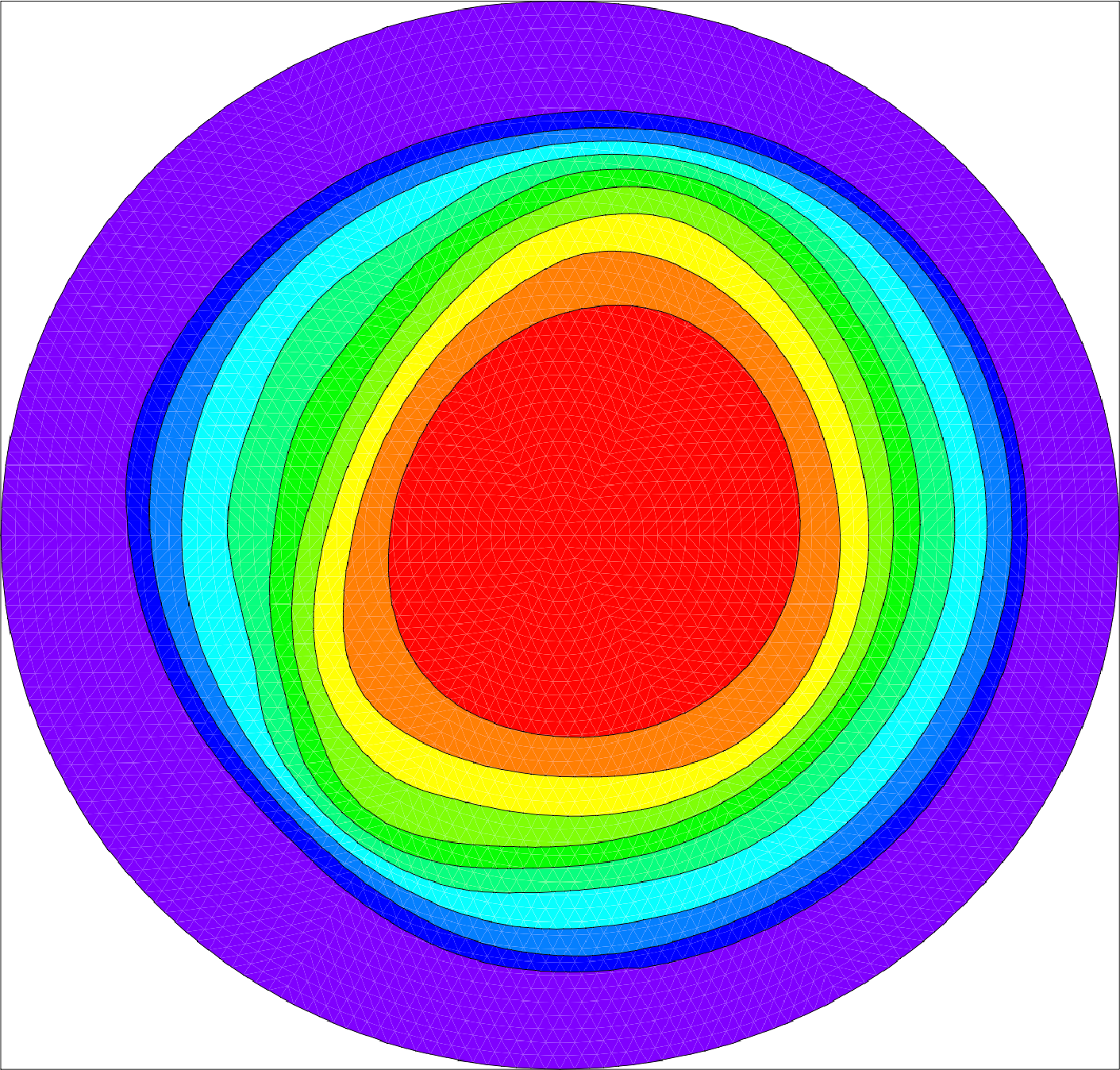}(a)
\includegraphics[height=2.5cm]{\figdir/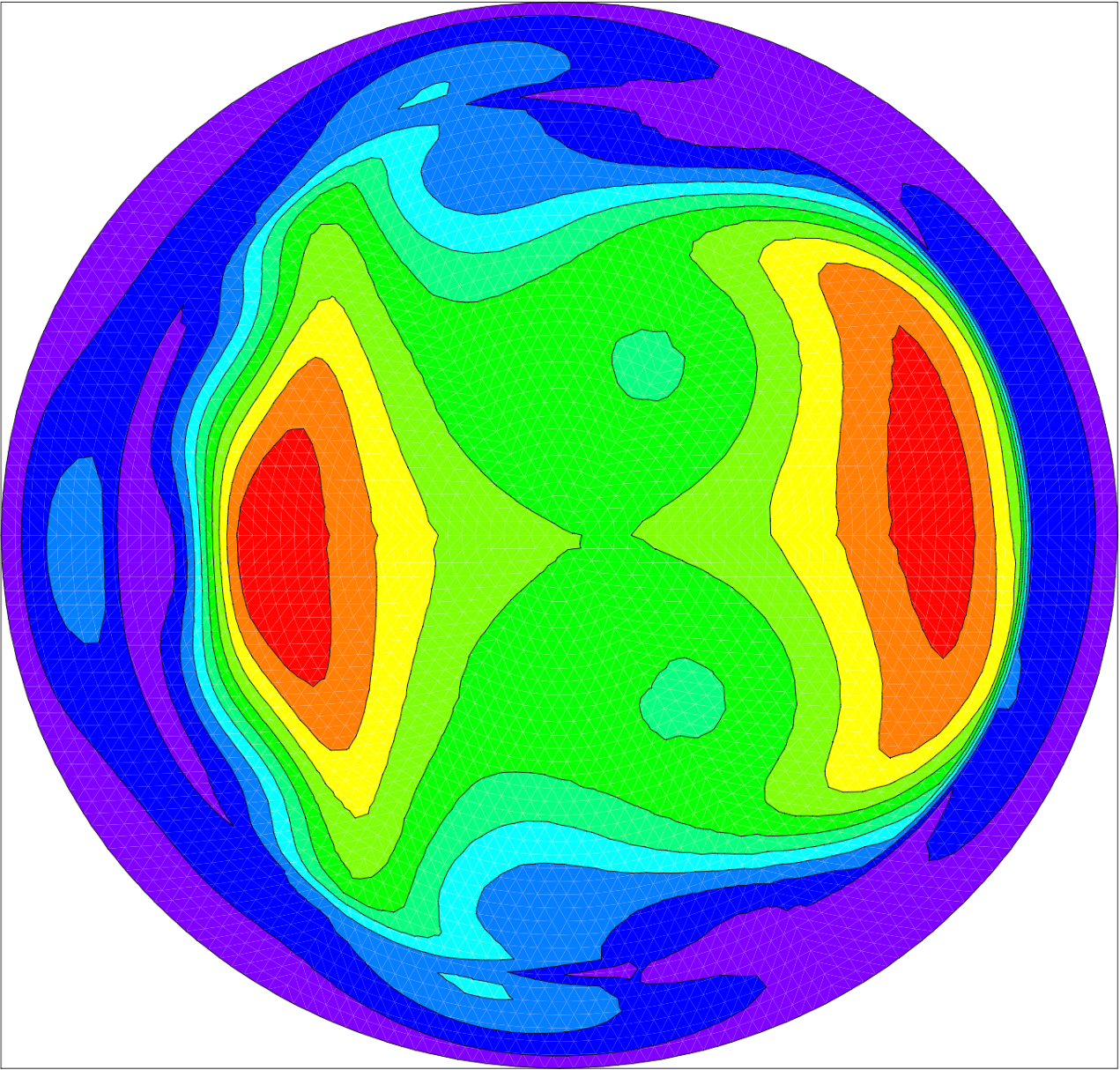}(b)
\includegraphics[height=2.5cm]{\figdir/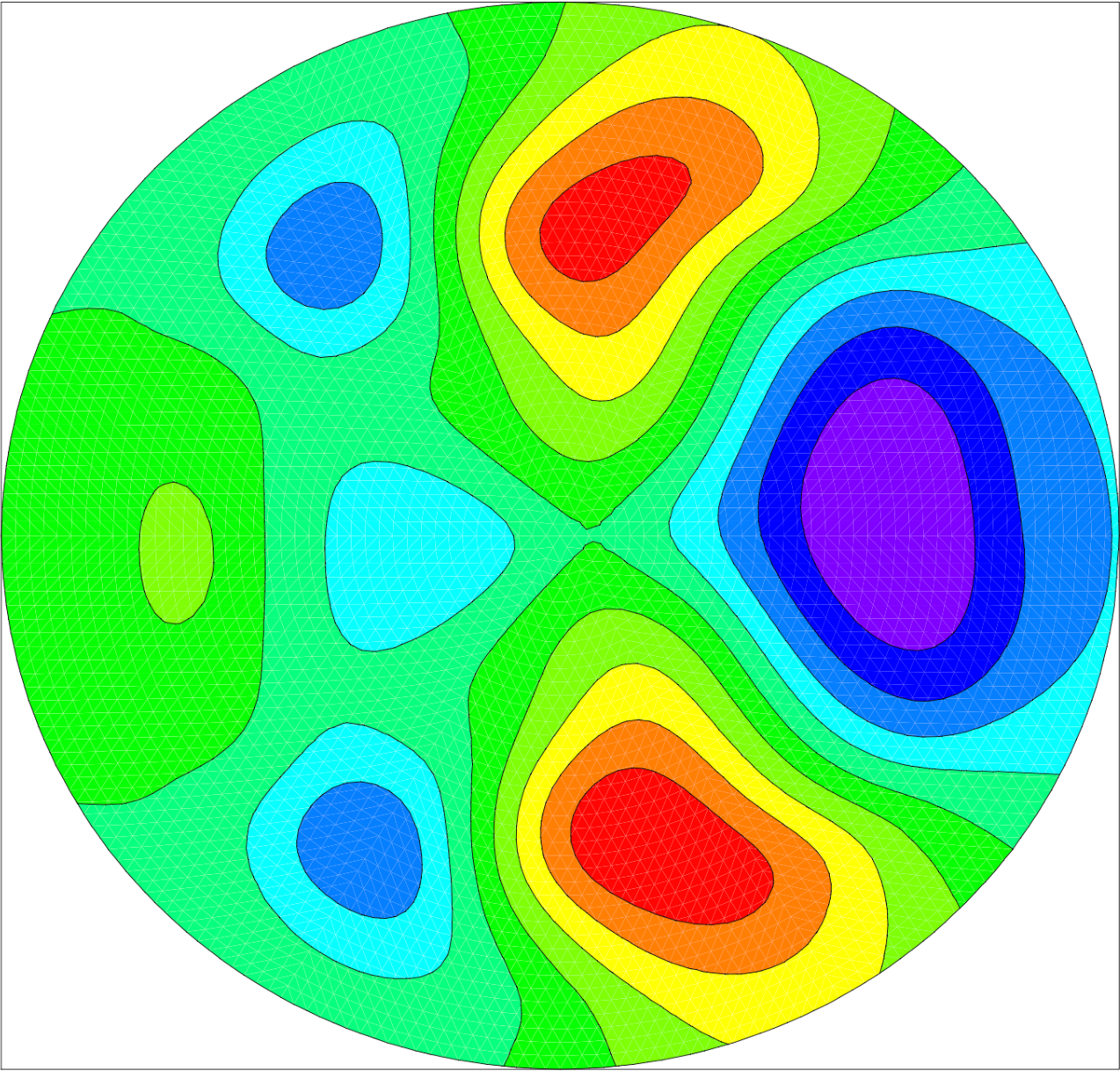}(c)
\includegraphics[height=2.5cm]{\figdir/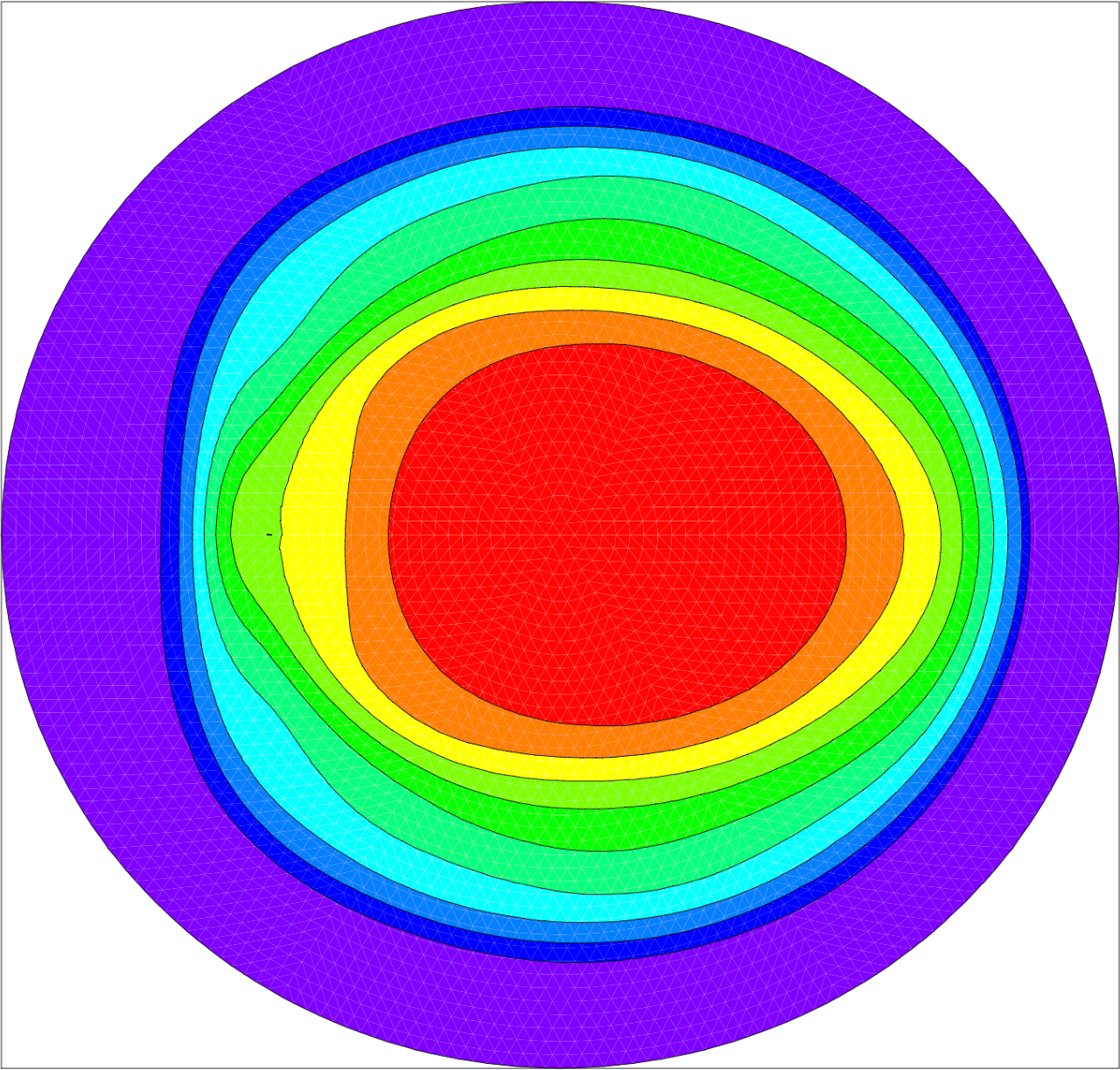}(d)
\includegraphics[height=2.5cm]{\figdir/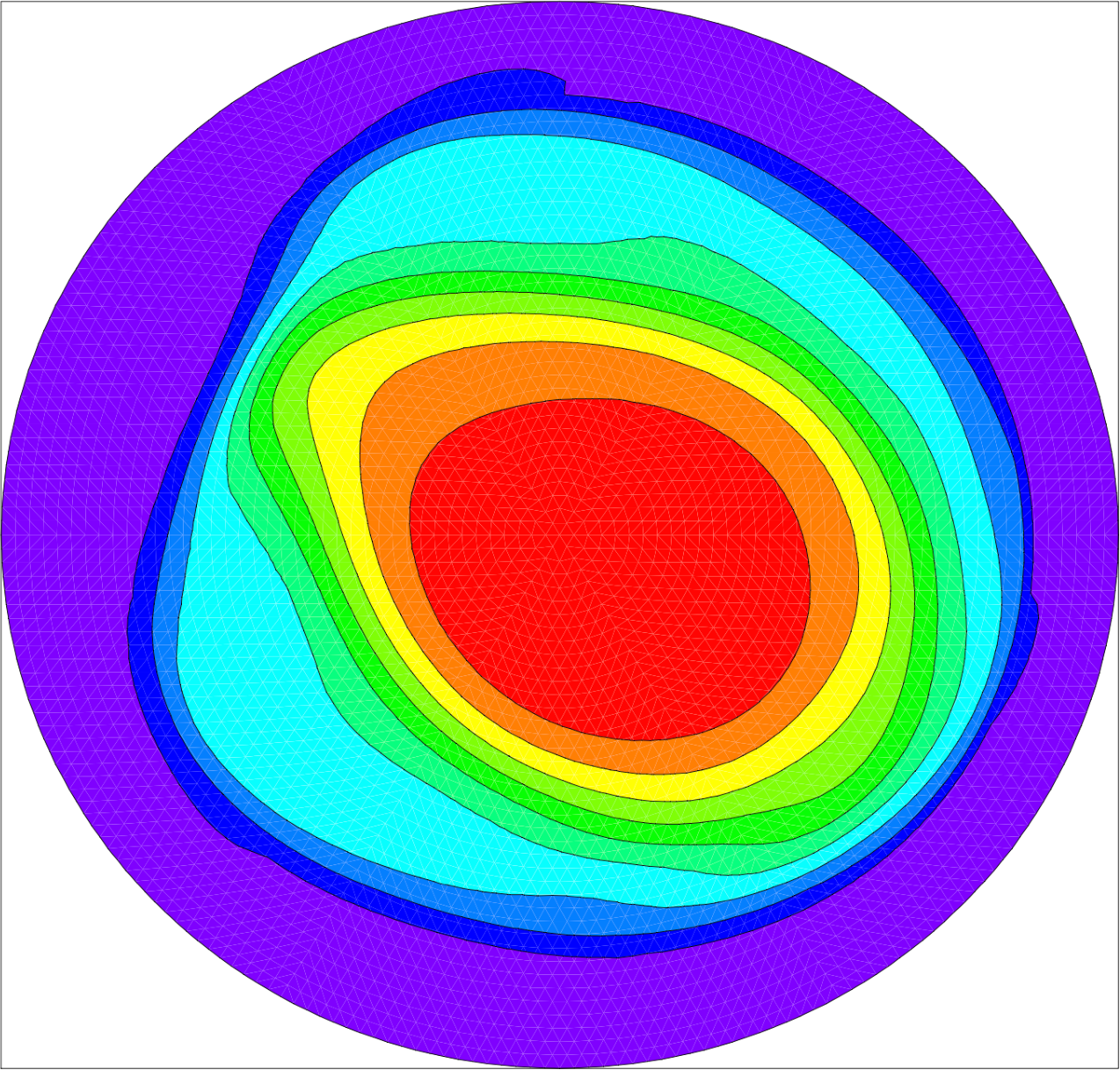}(e)
\end{center}
\caption{\it
Pressure $p$  contours  in nonlinear simulation of the $q_a$ case for 
(a) ideal wall,  
 (b) resistive wall,
(c) perturbed $n \ge 1$ part of magnetic field corresponding to  case (b),
(d) feedback stabilization,
(e) rotating wall. \rfig{lowbeta}(a),(b) reproduced from \cite{nf-iaea24}}
\label{fig:lowbeta}
\end{figure}
\rfig{lowbeta} shows contours of pressure and perturbed magnetic field in 
 nonlinear simulations corresponding to the time histories in \rfig{qmst}, 
 including feedback and
wall rotation.
The simulations demonstrate that
ideal wall, feedback, and rotating wall  limit growth of tearing mode. 
A resistive wall (or no wall)   allows a tearing mode  to 
reach much larger
amplitude than an ideal wall, or similar boundary conditions. 
Contours of pressure $p$ are shown near the last times in the history plot
\rfig{qmst}. The contour plots correspond to boundary conditions
(a) ideal wall, (b) resistive wall with no rotation, (d) feedback, and
(e) edge rotation. Boundary conditions (d) and (e) are similar to (a), with
small perturbations and only minor disruptions. In (c), perturbed magnetic flux $\psi$ 
contours correspond to the pressure contours in (b).
The perturbed flux is relevant  to the anaysis of $\rho_{q2}$ in \rsec{q75}.
  The feedback and wall rotation is  described  in \cite{nf-iaea24} and
summarized in the following \rsec{feedback}.

\section{Feedback} \label{sec:feedback}

Active feedback and wall rotation  can
make the wall effectively ideal and suppress RWTM major disruptions. 
There have been extensive theoretical  \cite{
bondeson-liu,he,brennan} and experimental studies of feedback stabilization
\cite{garofalo,okabayashi2009,sabbagh2010}.
To model feedback, consider the magnetic diffusion equation at a thin
resistive wall \cite{jet21,d3d22,finn95}
\be 
\frac{\pd \psi_w}{\pd t} 
= \frac{\eta_w}{\delta_w} ( \psi'_{vac} - \psi_p' ) - \Omega_w \frac{\pd \psi }{\pd \phi}.  \label{eq:thin0} \ee
where $\psi_w$ 
is the magnetic potential at the wall,
$\psi_p'$ is its radial derivative on the plasma side of the wall,
$\eta_w,\delta_w$ are the wall resistivity and thickness,
and $\psi'_{vac}$ is the radial derivative of $\psi_w$ on the vacuum
side of the wall. 
The vacuum field is taken of the form
\be \psi_{vac} = \psi_w \left(\frac{r_w}{r}\right)^m
  + \psi_f \left[ \left(\frac{r_w}{r}\right)^m - \left(\frac{r}{r_w}\right)^m \right] 
\label{eq:psivac} \ee
where $ \psi_f = g D\psi_w/2 - hr_w F\psi_p'/(2m)$ is the feedback signal, $g$ is the normal gain,
$h$ is the transverse gain,
$D(\theta,\psi_w), F(\theta,\psi_w)$ are screening functions of poloidal and
toroidal angle of the wall, modeling  the location of the sensors,
and $r_w$ is the wall radius.
For now, take $D = F =1.$
They could  be taken non zero in future numerical studies,
and might affect detailed predictions of the modeling.
The  $g$ term models  saddle coils which sense $b_n \propto \psi_w$, 
while $h$ models  probes which sense transverse perturbed magnetic field $b_l \propto \psi_p'$. 
The $\Omega_w$ term models a rotating wall boundary condition  \cite{okabayashi2,brennan}.
Wall rotation can provide electromagnetic torque to sustain sheared rotation
\cite{okabayashi2}.

Then \req{thin0},\req{psivac}  can be expressed
\be \frac{\pd \psi_w}{\pd t} = -\frac{m }{\tau_{wall}} [  
(1 - h) \psi_p' +  (1+ g) \psi_w / r_w ] 
- \Omega_w \frac{\pd \psi }{\pd \phi}. 
\label{eq:thin1} \ee
In the simulations in this paper, only $h$ or $\Omega_w$ are used, and are  constant in time. 
More advanced experimental methods vary the feedback gain in time \cite{sabbagh2010}.
The goal here is to demonstrate that feedback or wall rotation can prevent
major disruptions, although it can permit minor disruptions.

\section{High $\beta$ NSTX RWTM} \label{sec:nstx}
RWM and RWTM can be found together at high $\beta.$
Both can be feedback stabilized.
\begin{figure}[h]
\begin{center}
\includegraphics[width=6.85cm]{\figdir/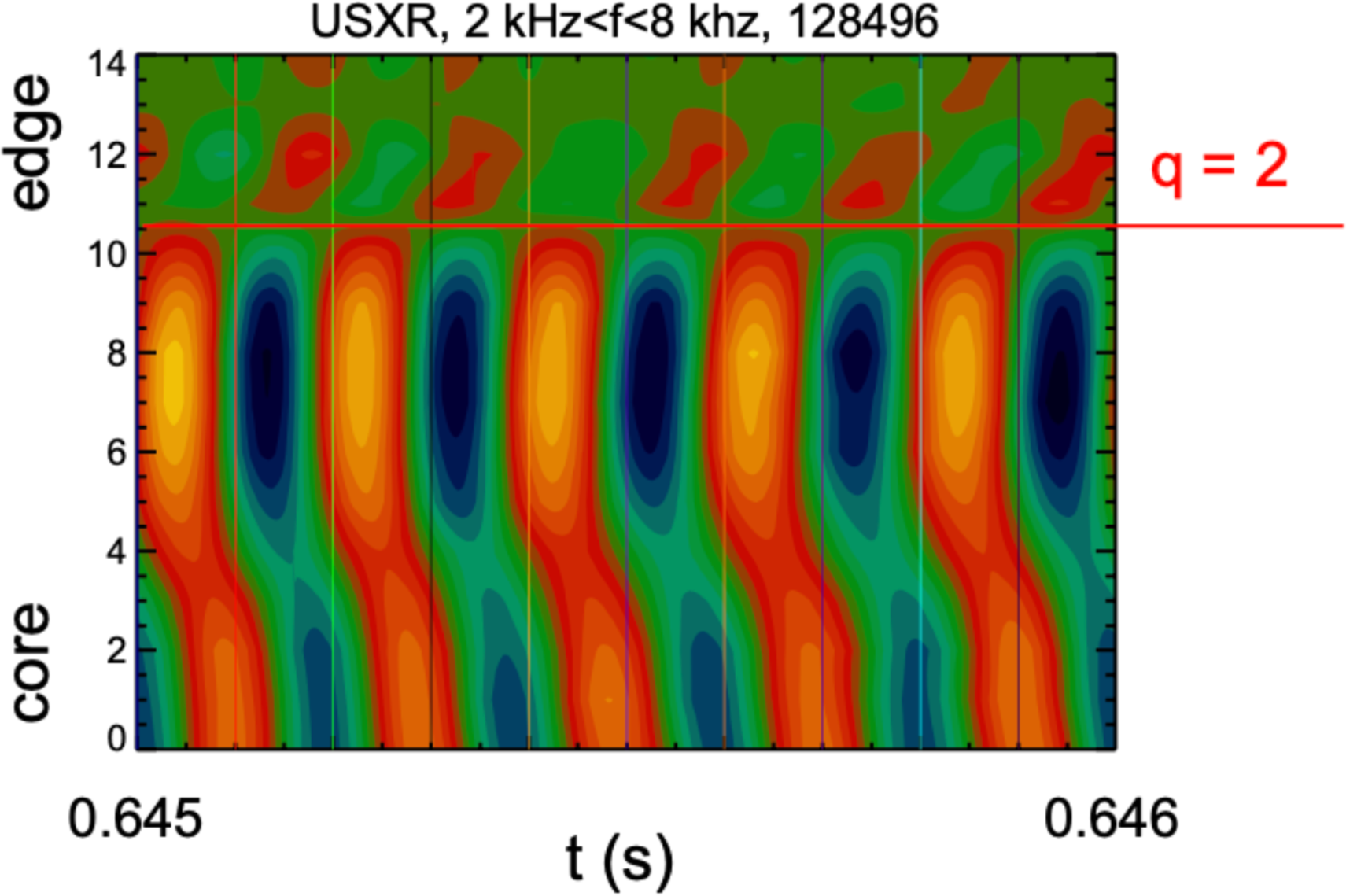} 
\end{center}
\caption{ \it  
feedback stabilized  $(2,1)$ RWTM. The RWTM can be identified by
its  phase inversion at  $\rho_{q2} = 0.75.$ Reproduced from \cite{sabbagh2010}.}
\label{fig:sabbagh} 
\end{figure}
\rfig{sabbagh} gives an
NSTX example \cite{sabbagh2010} , with $\beta_N > 4,$ above the no wall limit.  
The feedback is with complex gain, which can vary in time as the modes grow.
Time dependent soft X ray data  shows radial mode structure.
Initially 
a locked RWM is stabilized by feedback. 
It then spins up and  converts to a stabilized  external kink.
It then  becomes in \rfig{sabbagh} a feedback stabilized  $(2,1)$ RWTM. 
The RWTM can be identified by 
its  phase inversion in soft X ray emissiom  at  $\rho_{q2} \approx 0.75.$
It is a RWTM  because
it is close enough to the wall to be affected by
the feedback imposed  at the wall.
This suggests that initially 
$q_0 > 1$ on axis and  $\rho_{q2} < 0.75.$ 
Resistive evolution causes    current profile
peaking and decreases  $q_0$ on axis, pushing
$\rho_{q2} \ge 0.75.$ 
An example is seen in the simulations of \rsec{nstxsim}.

A similar phenomenon is seen was seen in DIII-D \cite{okabayashi2009}.
After an ELM, a  tearing mode developed in shot 131753  with 
$\rho_{q2} \approx   0.75,$ with growth time $10 ms,$ when the toroidal velocity 
 at $q_{75} \approx 0.$
The mode caused a major disruption, in which the ratio of initial to final $\beta$
dropped from $1$ to less than $0.25.$ 

It appears that RWTMs were observed in KSTAR \cite{kstar}. A simulation based on
an equilibrium reconstruction with $\beta_N = 3.7$ found a 
$(2,1)$ mode with magnetic perturbations extending through a resistive wall, 
clearly a RWTM although not identified as such.


\section{High $\beta$ NSTX  simulations  } \label{sec:nstxsim}
\begin{figure}[h]
\begin{center}
\includegraphics[height= 3.80cm]{\figdir/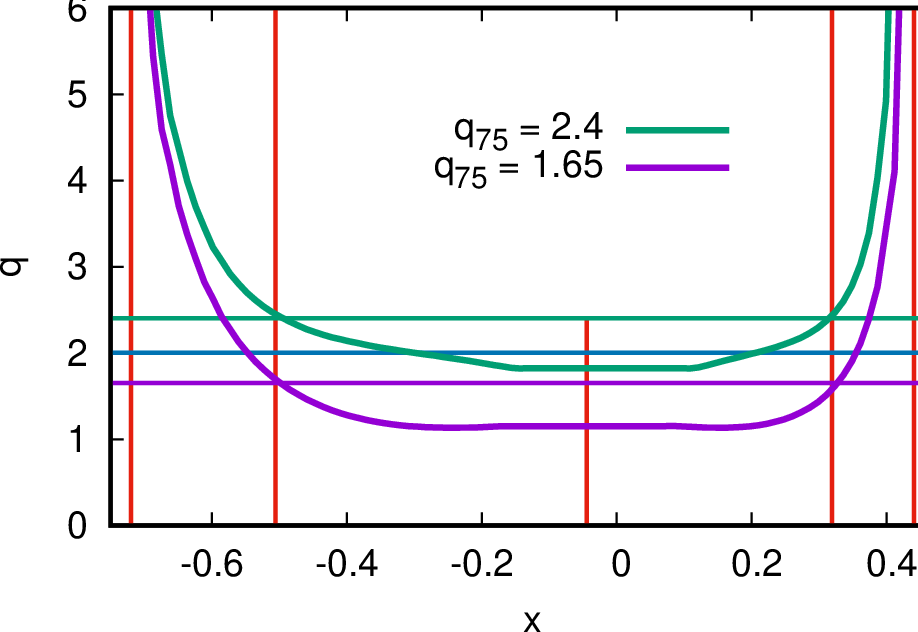}(a)
\includegraphics[height= 3.80cm]{\figdir/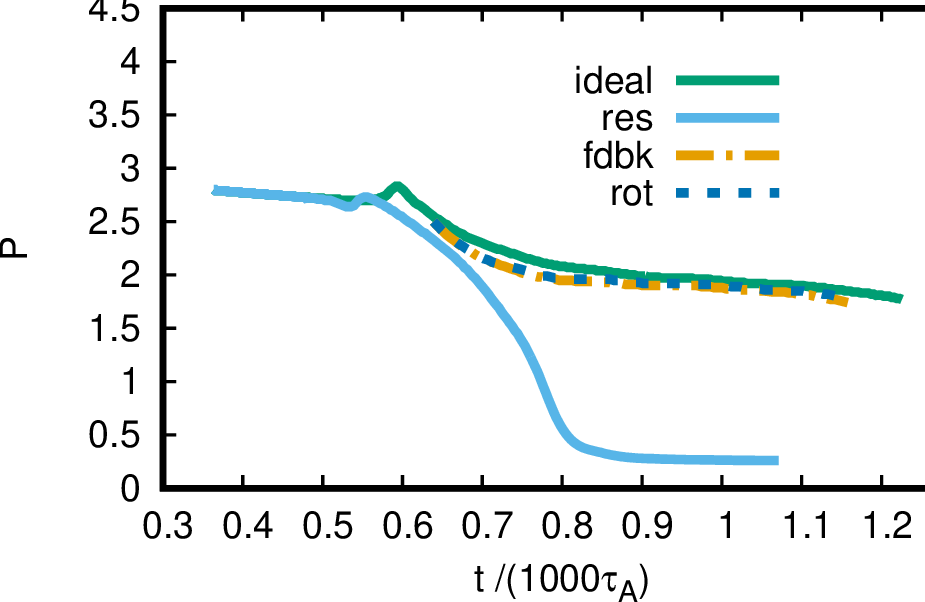}(b) 
\end{center}
\caption{\it (a) $q(x)$ profiles with evolution to RWTM instability.
(b) Time histories of total pressure $P$ with different  boundary conditions:
ideal wall, resistive wall, feedback, and rotating wall.}
\label{fig:nstx}
\end{figure}

Simulations were done with M3D of modified NSTX equilibrium reconstructions
of shot 109070.  The simulation parameters were the same as in \rsec{lowbeta}. 
An example is given in
\rfig{nstx} 
  with $\beta_N = 3$.
\begin{figure}[h]
\begin{center}
\includegraphics[width=2.4cm]{\figdir/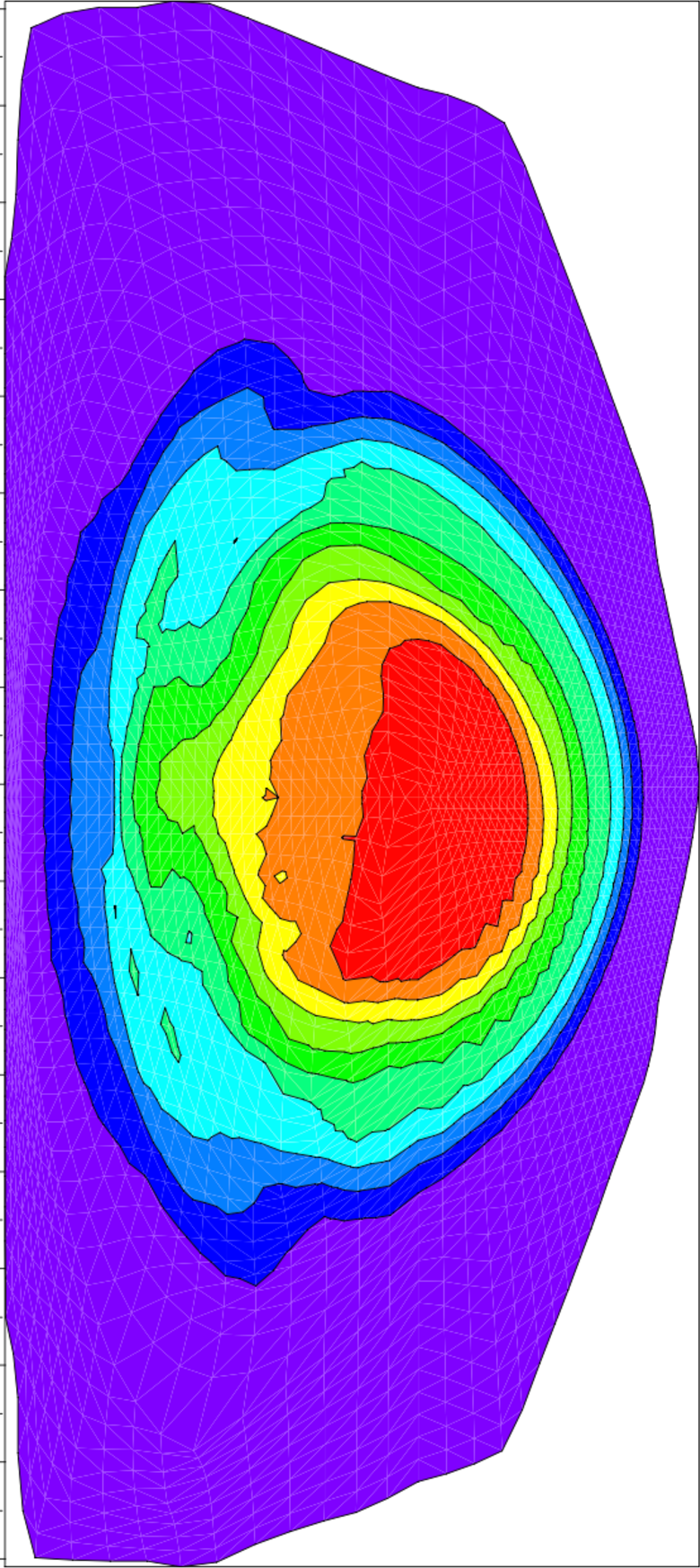}(a) 
\includegraphics[width=2.4cm]{\figdir/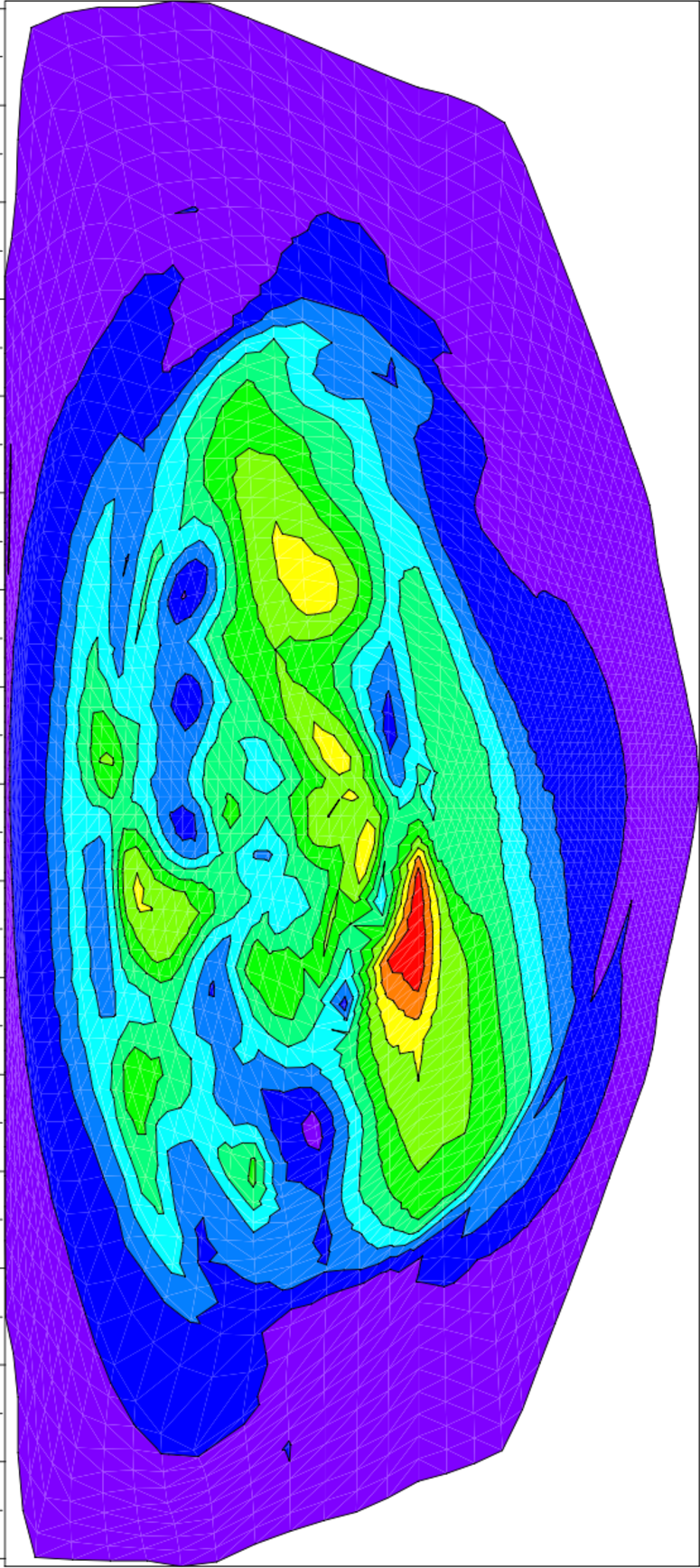}(b)
\includegraphics[width=2.4cm]{\figdir/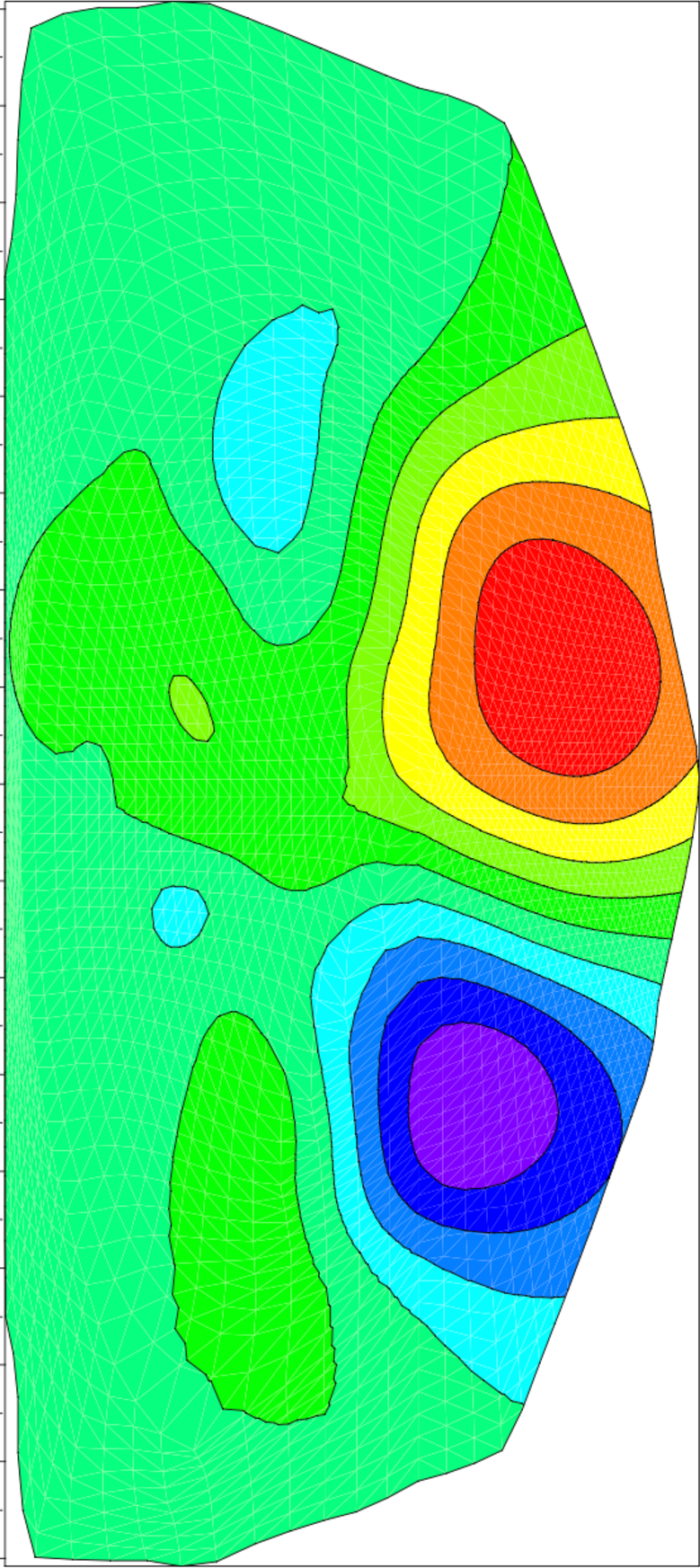}(c)
\includegraphics[width=2.4cm]{\figdir/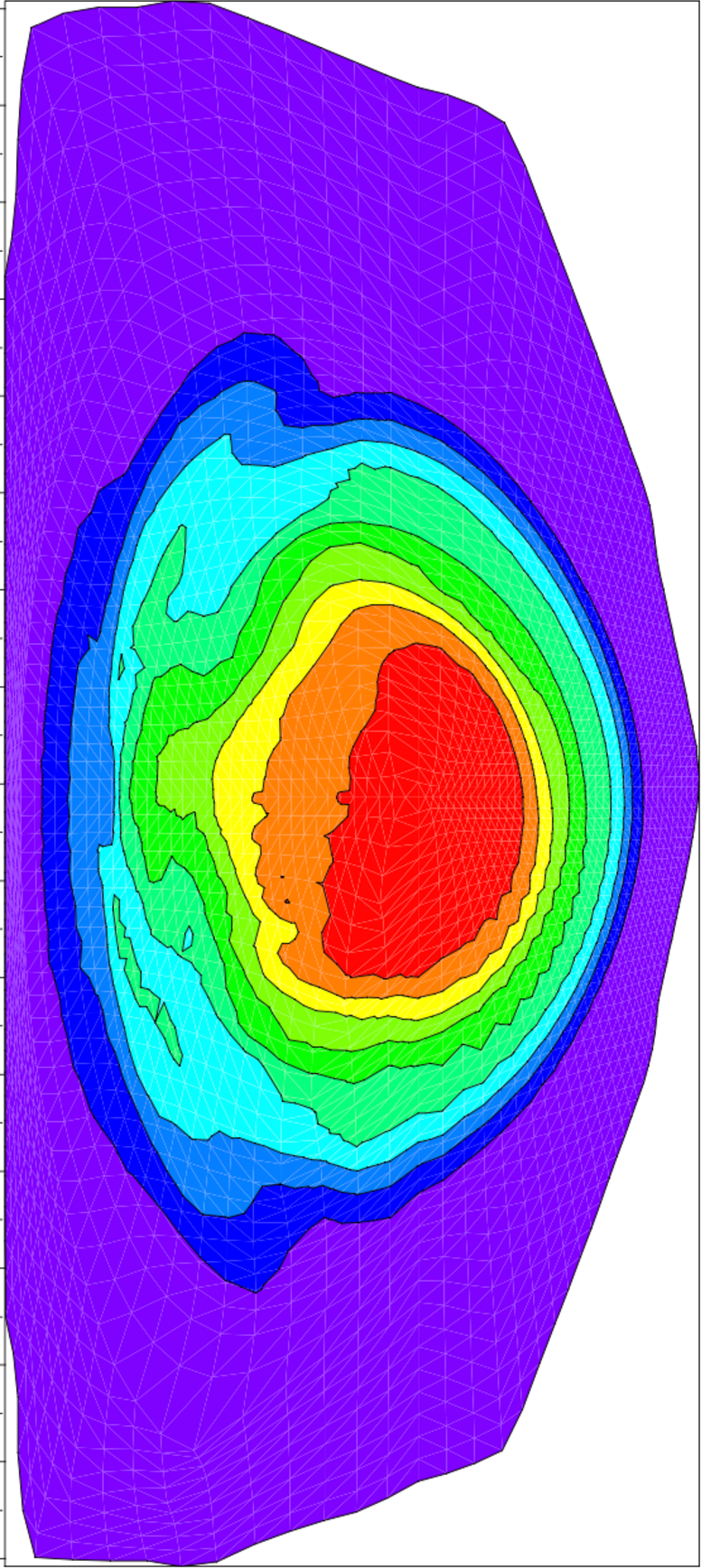}(d)
\includegraphics[width=2.4cm]{\figdir/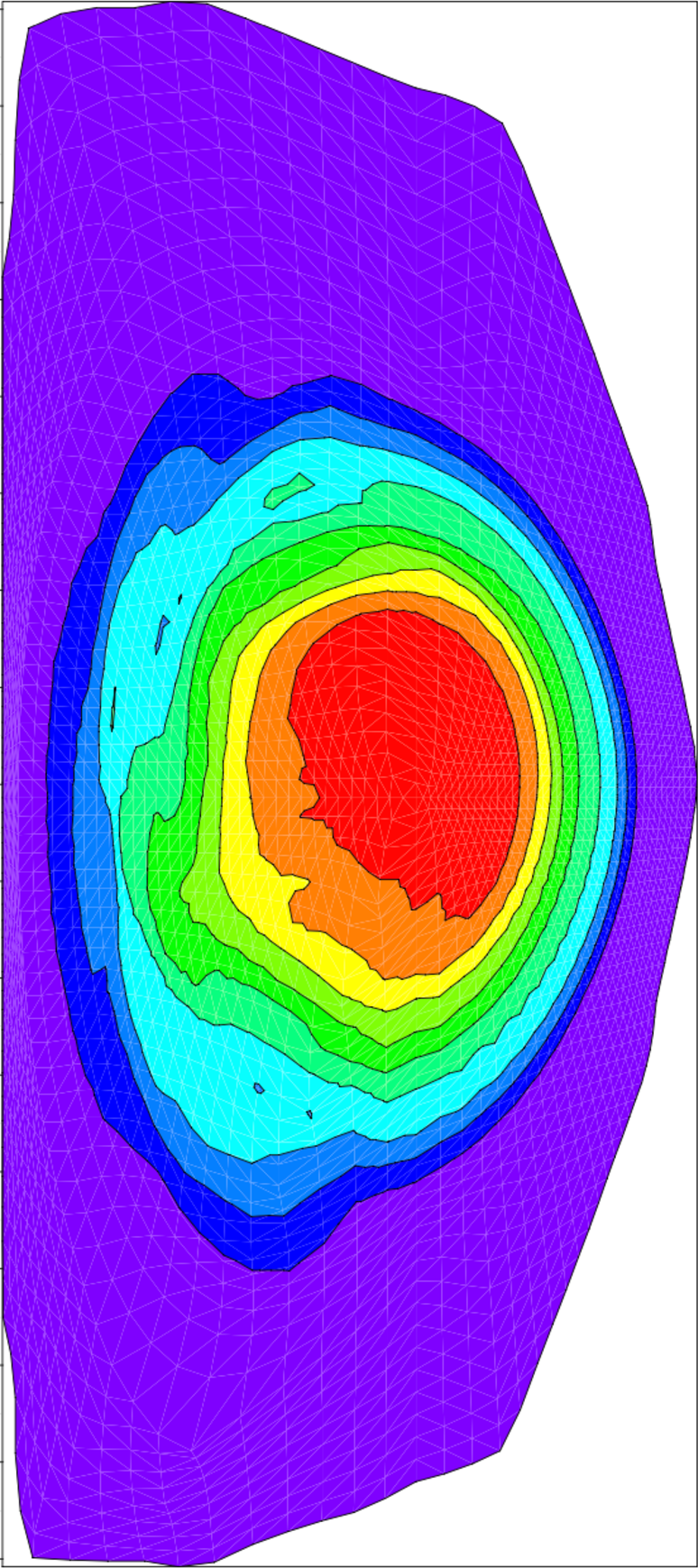}(e)
\end{center}
\caption{\it Contours of pressure  
near the end of the time histories in \rfig{nstx}(b).
with (a) ideal wall; (b) resistive wall; (c) perturbed magnetic flux of resistive case (b);
(d)  magnetic feedback; (e) rotating wall.}
\label{fig:nstx3n}
\end{figure}
\rfig{nstx}(a) gives midplane $q(x)$ profiles, $x = R - R_0,$ 
 at nearly the initial time,  when $q_0 = 1.3$ and  
$q_{75} =2.4$. It appears stable to a $(2,1)$ and $(3,1)$ mode. 
The equilibrium is allowed to evolve resistively, with the current contracting until  $q_0 \approx  1,$ and  
 $q_{75} = 1.65.$ The equilibrium is then  RWTM unstable to a $(2,1)$ mode. 
\rfig{nstx} (b) shows time histories of total pressure $P$  with different  boundary conditions:
ideal wall, resistive wall, feedback, and wall rotation. Only the case with
resistive wall without feedback or wall rotation has a major disruption. The
other cases all have minor disruptions.
\rfig{nstx3n} shows 
contours of pressure  
with (a) ideal wall; (b) resistive wall; (d)
feedback; (e) rotating wall. A major disruption occurs only with a locked resistive wall.
The pressure contours have a large perturbation, as in \rfig{lowbeta}.
\rfig{nstx3n}(c) shows $n > 1$ contours of $\psi.$
The perturbations are large lobes which penetrate the wall.

At high $\beta$, there are also RWMs, resistive wall external kink modes. Evidently
they can also be stabilized by feedback \cite{sabbagh2010} as mentioned in
\rsec{nstx}. Simulations of RWMs will be presented elsewhere.

\section{ Dependence of $\rho_{q2}$ on  wall position } \label{sec:q75}

\begin{figure}[h]
\begin{center}
 \includegraphics[width=6.0cm]{\figdir/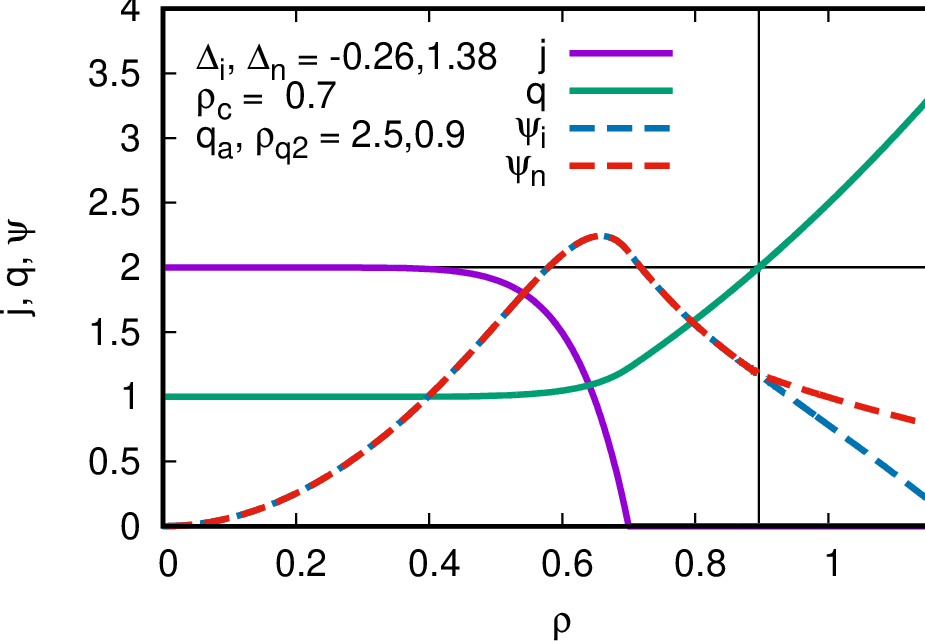}(a)
 \includegraphics[width=6.1cm]{\figdir/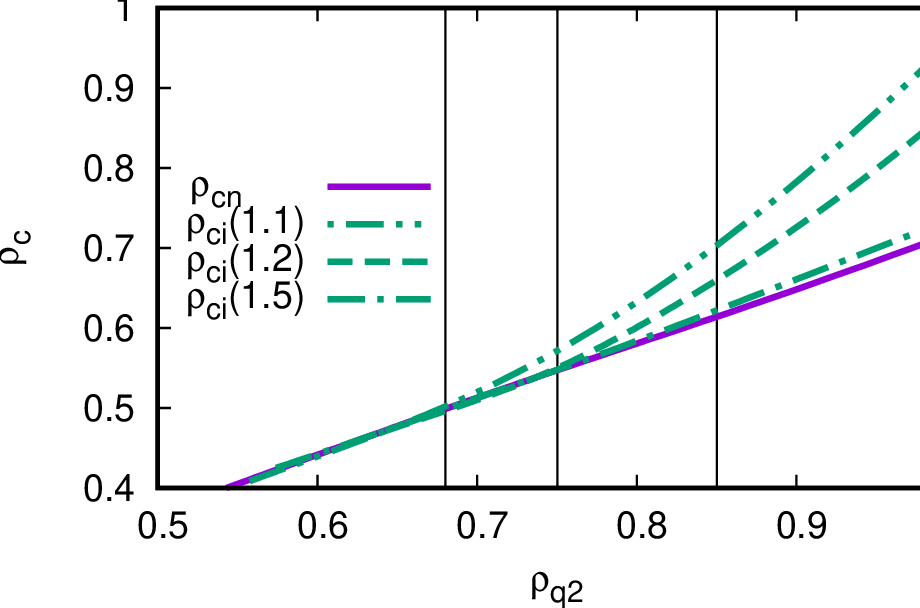}(b)
\end{center}
\caption{\it (a)  $\psi,$ $j$, and $q,$  with $\psi$ for ideal $(\psi_1)$  and no wall
$(\psi_2)$. (b)  Curves of $\rho_{ci}(\rho_{q2})$, 
$\rho_{cn}(\rho_{q2})$  and $q_a(\rho_{q2})$ 
for $\rho_{w} = 1.1, 1.2, 1.5.$} 
\label{fig:linear}
\end{figure}

The critical $\rho_{q2}$ depends on normalized wall radius $\rho_w.$ 
The critical value  $\rho_{q2} = .75$ occurs for
$\rho_w = 1.2,$ as in DIII-D, NSTX, and the MST model in \rsec{lowbeta}.
This can be obtained from  a linear model 
 \cite{model} using  modified \cite{frs73} equilibrium profiles
with current density $j(\rho) = 0$ for  $\rho > \rho_c,$ with
$ j(\rho) = (2/q_0) (1 + \rho^{2\nu})^{-(1 + 1/\nu)} - c_r $
with  $c_r = (1 + \rho_c^{2\nu})^{-(1 + 1/\nu)}$, and
 $q(0) = 1.$ 
The profile peakedness parameter $\nu$ is determined by $\rho_c$ and $q_a.$ 
Linear ideal MHD equations for perturbed magnetic flux $\psi$ with mode number $(2,1)$ 
were solved in a periodic
cylinder.    An example is given in \rfig{linear}(a), with $q_a = 2.5,$ $\rho_c = 0.7.$ 
The normalized $q=2$ radius is $\rho_{q2} = 0.9.$ Solutions of $\psi(\rho)$ are given
for an ideal wall boundary condition $\psi(\rho_w) = 0$ and a no wall boundary
condition $d \psi(\rho_w) / d \rho = -2 \psi(\rho_w) / \rho_w.$ 
The stability parameter 
$\Delta' = [\psi'(\rho_{q2+})-\psi'(\rho_{q2-})] /\psi(\rho_{q2})$
is calculated at $\rho_{q2}$ for ideal and no wall boundary conditions.
For an ideal wall, $\Delta' = -0.26,$ while for no wall, $\Delta' = 1.38.$ This is an
unstable RWTM. 
In \rfig{linear}(b) are plotted curves  
$\rho_{ci}(\rho_{q2},\rho_w)$ for $\Delta' = 0$
with ideal wall and $\rho_{cn}(\rho_{q2})$ with  $\Delta' = 0$ for no wall.
Three 
$\rho_{ci}$ 
curves are plotted, for $\rho_w = 1.1, 1.2, 1.5.$ 
There is only one $\rho_{cn}(\rho_{q2})$ curve,
since it does not depend on $\rho_{wall}.$
 The RWTM is unstable for $\rho_{ci} \ge \rho_c \ge \rho_{cn}.$
The onset condition for RWTM is $\rho_{ci} = \rho_{cn}.$
For $\rho_w = 1.2$,
$\rho_{q2} = 0.75$ as in \rfig{q75beta}(b). 
\begin{figure}[h]
\vspace{.5 cm}
\begin{center}
  \includegraphics[width=6.1cm]{\figdir/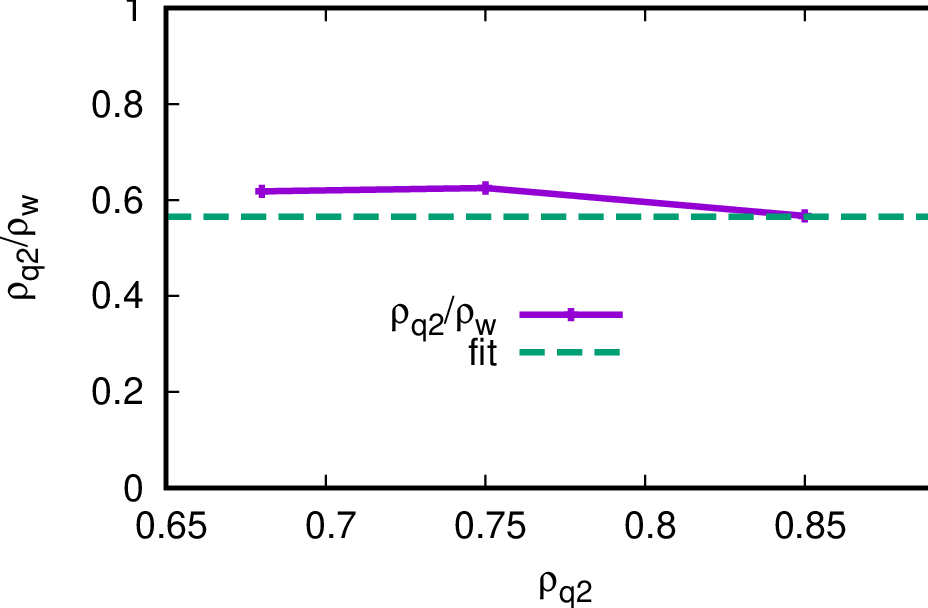}(a)
\includegraphics[width=3.9cm]{\figdir/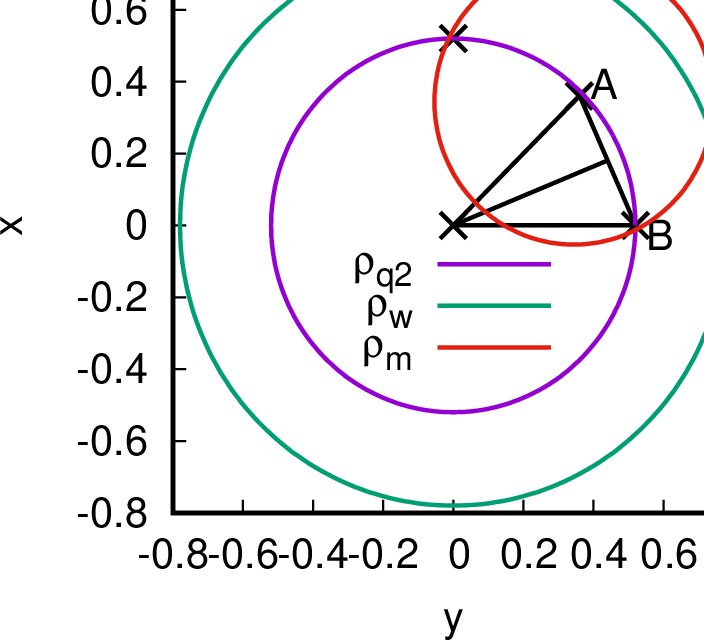} (b)
\end{center}
\caption{\it (a) $\rho_{q2}/\rho_w$ depends weakly on $\rho_{q2}$.
(b)  model of wall interaction of $(m,n) = (2,1)$ mode.}
\label{fig:linear2}
\end{figure}
\rfig{linear}(b)  gives a relation between $\rho_{q2}$ and $\rho_w,$ 
shown in \rfig{linear2}(a).
The value of $\rho_{q2}/\rho_w$
as a function of $\rho_{q2}$ is approximately constant. 
The data in \rfig{linear2}(a) can be fit as follows. 
The magnetic $n \ge 1$ perturbations shown in \rfig{lowbeta}(c) and
\rfig{nstx3n}(c) are lobes which extend into the wall.
In \rfig{linear2}(b), an $(m,n)$ mode is modeled as dividing the 
contour $\rho = \rho_{q2}$ into $2m$ arcs with ends at angles $0, \pi/m, \dots$
A chord of length $2 \sin(\pi / 2m) \rho_{q2}$  can be drawn connecting the midpoint of the arc
labelled ``A" to the intersection of the arc with the x axis at ``B". This is  
the radius $\rho_m$ of a  circle drawn through the midpoint of the arc ``A", 
as shown in \rfig{linear2}(b).
It  models  a lobe of a $(m,n)$  mode structure. 
The radius of the circle must be large enough to intersect the wall, 
such that  $\rho_{w} \le \rho_{q2} + \rho_{m}.$  
 This can be expressed 
\be \frac{\rho_w}{\rho_{q2}} \le 1 +  2 \sin( \frac{\pi}{4m}) \label{eq:rho} \ee  
which is $\rho_w/\rho_{q2} = 1.77$ for $m =2,$ $\rho_w/\rho_{q2} = 1.52$ for $m = 3.$
This is shown as the fit in \rfig{linear2}(a).
The calculated line in \rfig{linear2}(a) intersects   the  fit at $\rho_{q2} = 0.85,$ 
where according to \rfig{linear}(b), $\rho_w = 1.5.$ This suggests that for larger
$\rho_w,$ the wall is too far away to interact with the mode.


\section{ Summary  }\label{sec:summary}

To summarize,
resistive wall tearing modes (RWTM) can cause cause major disruptions.
A signature of RWTMs is that the rational surface is
sufficiently close to the wall. 
For $(m,n) = (2,1)$ modes, the  rational surface radius of the $q=2$ surface, 
normalized to the plasma minor radius, is $\rho_{q2} > 0.75.$
This can also be expressed as the value of $q$ at $\rho = 0.75$, 
$q_{75} < 2.$ 
The domain of instability of RWTMs
in the $(q_{75},\beta)$ plane was presented qualitatively.
The $\rho_{q2} > 0.75$ criterion was found in 
 a DIII-D locked mode disruption database.
The importance of mode locking and disruption precursors was  discussed.

A very important feature of RWTMs is that they produce major disruptions when the
$q_{75} < 2$ criterion is  satisfied. If this is not satisfied, or if the wall is ideally
conducting, then the mode becomes a tearing mode and does not produce a major disruption,
although it can produce a minor disruption. Feedback, or rotation of the
mode at the wall by complex feedback, can emulate an ideal wall. 
This was verified in simulations of a sequence of low $\beta$ equilibria. 
It was shown that when the wall is resistive and the $q_{75}$ criterion is
satisfied, the saturated mode amplitude is large, otherwise it is small. 
The computational model used for feedback and  rotating wall was discussed. 

At high $\beta$, feedback stabilized $(2,1)$ modes were observed in NSTX
with $\rho_{q2} \approx 0.75,$ indicating wall interaction, implying a RWTM.
In DIII-D, modes with $\rho_{q2} \approx 0.75$ caused a major disruption.
Simulations were performed using modified NSTX equilibria at moderate $\beta_N = 3,$
which satisfied $q_{75} < 2,$
and produced major disruptions with a resistive wall, minor disruptions with an
ideal wall, feedback, or rotating wall.

The $q_{75}$ criterion 
was analyzed in a linear simulations, and a
simple geometric model was given. 
The criterion depends weakly on the ratio of $\rho_{q2} / \rho_w,$ where
$\rho_w$ is the wall radius normalized to plasma radius.  For $\rho_w > 1.5,$ the
wall is too far away for a RWTM and feedback stabilization is not possible. The
criterion is also obtained for general $(m,n),$ and requires the rational surface
to be closer to the wall. 

In conclusion, $(m,n) = (2,1)$ RWTMs satisfy the $q_{55} < 2$ condition. 
The boundary conditions at the wall can
prevent a major  disruption.
With an ideally conducting wall, tearing modes produce only minor disruptions.
Feedback and rotating wall boundary conditions  act like an ideal wall.
This could potentially eliminate disruptions from tokamaks, greatly enhancing the
prospects of magnetic fusion.

{\bf Acknowledgement} Thanks to S. Sabbagh for pointing out the possible RWTM in
\cite{sabbagh2010}. This work was supported by  U.S. D.O.E. grant DE-SC0020127. 


\begin{thebibliography}{99}
\bibitem{jet21}
H. Strauss and JET Contributors,
Effect of Resistive Wall on Thermal Quench in JET Disruptions,
Phys. Plasmas \textbf{28}, 032501 (2021) 
\bibitem{iter21} H. Strauss, Thermal quench in ITER disruptions,
 Phys. Plasmas \textbf{28} 072507 (2021) 
\bibitem{d3d22} H. Strauss, B. C. Lyons, M. Knolker,
Locked mode disruptions in DIII-D and application to ITER,
Phys. Plasmas \textbf{29}  112508 (2022);
\bibitem{mst23} H. R. Strauss,  B. E. Chapman, N. C. Hurst,
MST Resistive Wall Tearing Mode Simulations, 
Plasma Phys. Control. Fusion \textbf{65}  084002 (2023).
\bibitem{model} H. R. Strauss,  Models of resistive wall tearing mode disruptions,
Phys. Plasmas 30, 112507 (2023); doi:10.1063/5.0172375
\bibitem{nf-iaea24} H. R.Strauss, B. E. Chapman, B. C. Lyons, Resistive Wall Tearing Mode Disruptions,
Nucl. Fusion \textbf{64} 106037 (2024); doi:10.1088/1741-4326/ad7272
\bibitem{sweeney} R. Sweeney, W. Choi, M. Austin,  
 M. Brookman, V. Izzo, M. Knolker, R.J. La Haye, 
A. Leonard, E. Strait, F.A. Volpe and The DIII-D Team,
Relationship between locked modes and thermal quenches in DIII-D,
Nucl. Fusion \textbf{58}, 056022 (2018)
\bibitem{sweeney2017} R. Sweeney, W. Choi, R. J. La Haye, S. Mao, K. E. J.
Olofsson, F. A. Volpe, and the DIII-D Team,
Statistical analysis of m/n = 2/1 locked and quasi - stationary modes
with rotating precursors in DIII-D, 
Nucl. Fusion 57 0160192 (2017). 
\bibitem{troyon}
F. Troyon, A.Roy, W.A.Cooper, F.Yasseen, A.Tumbull,
Beta limit in tokamaks: experimental and computational status,
Plasma Physics and Controlled Fusion textbf{30},  1597 (1988).
\bibitem{betti} R. Betti,
Beta limits for the n = 1 mode in rotating - toroidal - resistive plasmas
surrounded by a resistive wall, Phys. Plasmas 5, 3615 (1998).
\bibitem{villamora} H. R. Strauss, 
Linjin Zheng, M. Kotschenreuther, W.Park, S. Jardin, J. Breslau, A.Pletzer, R. Paccagnella, L. Sugiyama, 
M. Chu, M. Chance, A. Turnbull,  
Halo Current and Resistive Wall Simulations of ITER,
paper TH/2 - 2, 20th IAEA Fusion Energy Conference 2004, Villamora, Portugal (2004).
\bibitem{gerasimov2020}
S.N. Gerasimov,
 P. Abreu, G. Artaserse, M. Baruzzo, P. Buratti, I.S. Carvalho, I.H. Coffey,
E. De La Luna, T.C. Hender, R.B. Henriques, R. Felton, S. Jachmich,
U. Kruezi, P.J. Lomas, P. McCullen, M. Maslov, E. Matveeva, S. Moradi,
L. Piron1, F.G. Rimini, W. Schippers, C. Stuart, G. Szepesi, M. Tsalas,
D. Valcarcel, L.E. Zakharov and JET Contributors,
Overview of disruptions with JET-ILW,
Nucl. Fusion \textbf{60}  066028 (2020).
\bibitem{wang}
S. Wang,  Z. W. Ma, 
Influence of toroidal rotation on resistive tearing modes in tokamaks,
Phys. Plasmas 22, 122504 (2015); doi:10.1063/1.4936977
\bibitem{coelho} R. Coelho,  E. Lazzaro, 
Effect of sheared equilibrium plasma rotation on the classical tearing mode 
in a cylindrical geometry, 
Phys. Plasmas 14, 012101 (2007)
\bibitem{drift} H. R. Strauss,
Rotational stabilization of drift tearing modes,
Phys. Fluids \textbf{B 4} 3 (1992); doi:10.1063/1.860448
\bibitem{finn95} J. A. Finn,
 Resistive wall stabilization of kink and tearing modes, Phys. Plasmas 2, 198 (1995)
\bibitem{schuller}
F.C. Schuller, Disruptions in tokamaks, 
Plasma Phys. Controlled Fusion \textbf{37}, A135 (1995).
\bibitem{wesson}
J.A. Wesson, R.D. Gill, M. Hugon, F.C. Schuller, J.A. Snipes, D.J. Ward, D.V. Bartlett,
D.J. Campbell, P.A. Duperrex, A.W. Edwards, R.S. Granetz, N.A.O. Gottardi, T.C. Hender,
E. Lazzaro, P.J. Lomas, N. Lopes Cardozo, K. F. Mast,
M.F.F. Nave, N.A. Salmon, P. Smeulders, P.R. Thomas, B.J.D. Tubbing, M.F. Turner, A. Weller,
Disruptions in JET, Nucl. Fusion \textbf{29} 641 (1989).
\bibitem{ricci} 
M. Giacomin, A. Pau, P. Ricci, O. Sauter, T. Eich, the ASDEX Upgrade team, JET Contributors, and the TCV team,
First-Principles Density Limit Scaling in Tokamaks Based on Edge Turbulent Transport
and Implications for ITER 
Phys. Rev. Lett. 128, 185003 (2022)
\bibitem{pucella} G. Pucella, P. Buratti, E. Giovannozzi, E. Alessi,
F. Auriemma, D. Brunetti, D. R. Ferreira, M. Baruzzo,
D. Frigione, L. Garzotti, E. Joffrin, E. Lerche, P. J. Lomas, S. Nowak, L. Piron,
F. Rimini, C. Sozzi, D. Van Eester, and JET Contributors,
Tearing modes in plasma termination on JET:
the role of temperature hollowing and edge cooling, 
 Nucl. Fusion \textbf{61} 046020 (2021)
\bibitem{gates} D. A. Gates, D. P. Brennan, L. Delgado-Aparicio, and R. B. White,
The tokamak density limit: A thermo- resistive disruption mechanism,
Phys. Plasmas 22, 060701 (2015); https://doi.org/10.1063/1.4922472
\bibitem{izzo} V. A. Izzo, D. G. Whyte, R. S. Granetz, P. B. Parks,
E. M. Hollmann, L. L. Lao, J. C. Wesley, 
Magnetohydrodynamic simulations
of massive gas injection int Alcator C - Mod and DIII-D plasmas,
Phys. Plasmas \textbf{15}, 056109 (2008).
\bibitem{okabayashi2} M. Okabayashi,
P. Zanca, E.J. Strait, A.M. Garofalo, J.M. Hanson, Y. In5, R.J. La Haye, L. Marrelli, P. Martin, 
R. Paccagnella, C. Paz-Soldan,
P. Piovesan, C. Piron, L. Piron, D. Shiraki, F.A. Volpe
and The DIII-D and RFX-mod Teams,
Avoidance of tearing mode locking with electro-magnetic torque introduced by feedback-based 
mode rotation control in DIII-D and RFX-mod, Nuclear Fusion 57, 016035 (2017) 
\bibitem{m3d}  W. Park, E.  Belova, G. Y.   Fu,
X.  Tang, H. R.  Strauss, L. E.  Sugiyama,
Plasma Simulation Studies using Multilevel Physics Models,
  Phys. Plasmas 6, 1796 (1999).
\bibitem{bondeson-liu} A. Bondeson, Yueqiang Liu, D. Gregoratto, Y. Gribov and V.D. Pustovitov,
Active control of resistive wall modes in the large-aspect-ratio tokamak,
Nucl. Fusion 42 (2002) 768–779
\bibitem{he}
Yuling He, Yueqiang Liu, Xu Yang, Guoliang Xia, Li Li,
Active control of resistive wall mode via modification of external tearing index,
Physics of Plasmas 28, 012504 (2021)
\bibitem{brennan} D. P. Brennan, J. M. Finn,
Control of linear modes in cylindrical resistive magnetohydrodynamics with
a resistive wall, plasma rotation, and complex gain,
Phys. Plasmas \textbf{21}, 102507 (2014).
\bibitem{garofalo} 
A.M. Garofalo, G.L. Jackson, R.J. La Haye, M. Okabayashi, H. Reimerdes, E.J. Strait,
J.R. Ferron, R.J. Groebner, Y. In, M.J. Lanctot, G. Matsunaga, G.A. Navratil, W.M. Solomon,
H. Takahashi, M. Takechi, A.D. Turnbull and the DIII-D Team,
Stability and control of resistive wall modes in high beta, low rotation DIII-D plasmas,
Nucl. Fusion \textbf{47} 1121–1130 (2007).
\bibitem{okabayashi2009} 
M. Okabayashi, I.N. Bogatu, M.S. Chance, M.S. Chu, A.M. Garofalo, Y. In, G.L. Jackson, R.J. La Haye, 
  M.J. Lanctot, J. Manickam, L. Marrelli, P. Martin, G.A. Navratil, H. Reimerdes, E.J. Strait, 
H. Takahashi, A.S. Welander, T. Bolzonella, R.V. Budny, J.S. Kim,
R. Hatcher, Y.Q. Liu and T.C. Luce, 
Comprehensive control of resistive wall modes in DIII-D advanced tokamak plasmas,
Nucl. Fusion 49 (2009) 125003.
\bibitem{sabbagh2010} S. A. Sabbagh, S.P. Gerhardt, J.E. Menard, R. Betti, D.A. Gates, B. Hu, 
O.N.  Katsuro-Hopkins, B.P. LeBlanc, F.M. Levinton, J. Manickam, K. Tritz and H. Yuh,
Advances in global MHD mode stabilization research on NSTX,
Nucl. Fusion \textbf{50} 025020 (2010).
\bibitem{kstar}
Y.S. Park, S.A. Sabbagh, J.H. Ahn, B.H. Park, H.S. Kim, J.W. Berkery,
J.M. Bialek, Y. Jiang, J.G. Bak, A.H. Glasser, J.S. Kang, J. Lee, H.S. Han,
S.H. Hahn, Y.M. Jeon, J.G. Kwak, H.K. Park, Z.R. Wang, J.-K. Park,
N.M. Ferraro  and S.W. Yoon,
Analysis of MHD stability and active mode control on KSTAR for high
confinement, disruption-free plasma, Nucl. Fusion 60 056007 (2020); doi:10.1088/1741-4326/ab79ca
\bibitem{frs73} H. P. Furth, P. H. Rutherford, and H. Selberg,
 Tearing mode in the cylindrical tokamak,
Physics of Fluids 16, 1054 (1973)

\end{thebibliography}
\end{document}